\documentclass[12pt]{article}

\pdfoutput=1

\usepackage[dvipdfmx]{graphicx}
\usepackage{graphicx,bm,color}
\usepackage{amsmath}
\usepackage{amssymb}
\usepackage{amsfonts}
\usepackage{cases}
\usepackage{cancel}
\usepackage{hyperref}
\usepackage{ulem}
\usepackage{cite}
\usepackage{xcolor}
\usepackage{setspace}

\usepackage[a4paper,margin=27mm]{geometry}

\begin{document}

\vspace*{8mm}

\begin{center}

{\Large\bf  Yukawa coupling unification in non-supersymmetric}

\vspace*{2mm}

{\Large\bf  SO(10) models with an intermediate scale}

\vspace*{9mm}

\mbox{\sc Abdelhak Djouadi$^{1,2}$, Ruiwen Ouyang$^2$, Martti Raidal$^2$ }\vspace*{3mm}

{\small

$^1$ Centro Andaluz de Fisica de Particulas Elementales (CAFPE) \\ and \\ Departamento de F\'isica Te\'orica y del Cosmos, \\ Universidad de Granada, E--18071 Granada, Spain\\ \vspace{0.3cm}

$^2$ Laboratory of High Energy and Computational Physics, NICPB, R\"avala pst. 10,\\ 10143 Tallinn, Estonia\\ \vspace*{0.3cm}
}

\end{center}

\vspace{20pt}

\begin{abstract}

\noindent We discuss the possibility of unifying in a simple and economical manner the Yukawa couplings of third generation fermions in a non-supersymmetric SO(10) model with an intermediate symmetry breaking, focusing on two possible patterns with intermediate Pati-Salam and minimal left-right groups. For this purpose, we assume a minimal Yukawa sector at high energy, starting with two Higgs bi-doublets at the intermediate scale which then simply reduce to a two Higgs doublet model at the electroweak scale. We first enforce gauge coupling unification at the two-loop level by including the threshold corrections in the renormalisation group running which are generated by the heavy fields that appear at the intermediate symmetry breaking scale. We then study the running of the Yukawa couplings of the top quark, bottom quark and tau lepton at two-loops in these two breaking schemes, when the appropriate matching conditions are imposed. We find that the unification of the third family Yukawa couplings can be achieved while retaining a viable spectrum, provided that the ratio of the vacuum expectation values of the two Higgs doublet fields is large, $\tan\beta \approx 60$.  

\end{abstract}

\clearpage

\subsection*{1 Introduction}

The paradigm of grand unification \cite{Georgi,PS} is at the heart of particle physics as it exploits the power of symmetries to unify in a most elegant way the electromagnetic, weak and strong interactions of the Standard Model (SM) into a single force \cite{Weinberg}. Grand unified theories (GUTs)  provide natural solutions to theoretical questions such as charge quantization and anomaly cancellation, in addition to the explanation of the existence of three separate gauge symmetry groups. GUTs can also successfully address most, if not all, of the important issues that call for beyond the SM physics. This  is particularly the case for the problems of neutrino masses and mixing, the baryon asymmetry in the universe and the nature of the dark matter. Hence, leaving aside the issue of naturalness and the large hierarchy between the weak and Planck scales that induces quadratic ``divergences" to the observed Higgs boson mass (for which one can, for instance, adopt an anthropic point of view just as in the case of  the cosmological constant), non-supersymmetric GUTs can be viewed as the royal path to physics beyond the SM. 

 Unification in the context of SO(10) \cite{SO10} is particularly interesting as this symmetry group possesses a representation of dimension 16 in which, for each generation, one can accommodate the 15 chiral fermions of the SM and an additional Majorana neutrino. If the mass of this new state is very large, somewhere at a scale  of $10^{10}$ GeV,  the see-saw mechanism~\cite{see-saw} could explain the present pattern in the neutrino sector,  the baryon asymmetry could be achieved through leptogenesis~\cite{baryogenesis} and a suitable axion~\cite{axion} could account for dark matter; see Refs.~\cite{Babu,Bajc,Altarelli} for reviews. This intermediate scale can naturally be present in SO(10) as the group is of rank five, {\it i.e.,} larger than the rank of the SM group by one unit, so that the symmetry breaking may occur in three steps, one at the GUT scale $M_U$, one at this intermediate scale $M_I$ and a last one at the electroweak scale. This solves one of the main drawbacks of non-supersymmetric GUTs, namely, the failure of the gauge couplings to unify at the high energy scale. Indeed, threshold effects \cite{Threshold} are generated by the contributions of the scalar multiplets that break the intermediate symmetry down to the SM group at the energy $M_I$, and these modify the renormalisation group evolution of the three coupling constants such that they finally intersect at the scale $M_U$~\cite{Unif,Meloni:2019jcf}. Hence, gauge coupling unification can also be realized without the need of supersymmetry, which was one of its main attractive points~\cite{SUSY-Unif}. 
 
 Another argument in favor of supersymmetry was the possibility of also unifying the Yukawa couplings of third generation fermions \cite{SUSY-YUC}. Indeed, in the minimal supersymmetric extension of the SM, the MSSM,  two-Higgs doublets fields are required in order to generate separately masses for the isospin up- and down-type  fermions and, in constrained scenarios with universal ``soft" SUSY--breaking parameters, these Yukawa couplings can be unified at the GUT scale. This occurs at large values of the ratio of the vacuum expectation values of the two Higgs fields, $\tan\beta=v_u/v_d$, which induces the proper hierarchy for the starting top and bottom quark masses, $\tan\beta \approx m_t/m_b  \approx 60$.
 
In this letter, we show that the unification of the Yukawa couplings of third generation fermions can also be achieved in a rather simple and most economical way in a non-supersymmetric SO(10) scenario taking as examples two of the most interesting and widely discussed intermediate breaking patterns: the Pati-Salam \cite{PS} and the minimal left-right symmetric \cite{LR} groups. As a matter of fact, and in contrast to most earlier studies, only two Higgs bi-doublet fields will be necessary to describe the Yukawa interactions of standard fermions above the scale at which the intermediate breaking occurs, and this spectrum then reduces to two Higgs doublets only below this intermediate scale and down to the electroweak scale. Hence, one would have an effective two Higgs doublet model (2HDM) of type II \cite{Branco:2011iw} at low energies, just as in the MSSM, with vacuum expectation values such that the parameter $\tan\beta$  is large as to obtain the correct hierarchy for the top and bottom quark masses. Using this minimal scalar sector, it is possible to make that the renormalisation group running of third generation Yukawa couplings in these two breaking schemes, with suitable matching conditions at the intermediate scale for which gauge coupling unification occurs, leads to Yukawa coupling unification  at the GUT scale. This can be achieved while reproducing the third family fermion and electroweak gauge boson masses and preserving some important features such as  ensuring the stability of the electroweak vacuum up to the intermediate scale and keeping the Yukawa couplings perturbative at all scales.

The paper is organized as follows. In the next section, we introduce our theoretical framework and discuss the breaking of SO(10) with intermediate steps. In section 3, we discuss the known issue of gauge couplings unification in SO(10) when threshold corrections are added at an intermediate scale but with a new ingredient, namely, the presence of an additional Higgs doublet field at low energies.  In section 4, we study the running of the third generation Yukawa couplings and show that they can reach a common value at the same scale  that allows for gauge coupling unification, while keeping a viable low energy spectrum. Our conclusions are given in section 5.
  
\subsection*{2 Theoretical framework}

The SO(10) group  has many attractive features \cite{Babu,Bajc,Altarelli} and most of them follow from the fact that it possesses a fundamental representation of dimension-16 in which, for each generation, the 15 SM chiral fermions as well as one right-handed neutrino can be embedded. In this case, the Yukawa couplings of the scalar bosons to pairs of these fermions belong to the direct product of $\mathbf{16 \otimes 16}$, which can be decomposed into 
\begin{eqnarray}
\mathbf{ 16_F \otimes 16_F} = \mathbf{10} + \mathbf{120} + \mathbf{126} \, .
\end{eqnarray}
Thus, the most general Yukawa interaction which is SO(10) invariant is given by
\begin{eqnarray}
-{\cal L}_Y = \mathbf{16_F} (Y_{10} \mathbf{10_H} + Y_{126} \mathbf { \overline{126}_H} + Y_{120} \mathbf{ 120_H} ) \mathbf{16_F} \, .
\label{eq:LY-general}
\end{eqnarray}
The special case with only the first two Yukawa terms with the $\mathbf{10_H}$ and $ \mathbf{\overline{126}_H}$ representations, leading to the so-called minimal SO(10) model, has been thoroughly discussed, see {\it e.g.} Refs.~\cite{Bajc,SO10-fits}. The extended SO(10) model including the $\mathbf{120_H}$ representation has been also explored \cite{Joshipura:2011nn,Babu:2016bmy}. The first model usually requires an extra U(1) symmetry to complexify the $\mathbf{10_H}$ representation to achieve the required splitting in the fermionic spectrum, otherwise the ratio $m_t/m_b$ would be fixed to unity at the GUT scale~\cite{Bajc}. In turn, in the latter scenario, it has been shown that a realistic fermion spectrum can be achieved with or without introducing such an extra U(1) symmetry~\cite{Babu:2016bmy}. 

In this work, we will restrict to the minimal and most studied SO(10) scenario in which only the $\mathbf{10_H}$ and $\mathbf{ \overline{126}_H}$ representations are kept, but without an extra U(1) symmetry to complexify the $\mathbf{10_H}$ representation. This will constrain the parameter space, making the model more predictive, while allowing the possibility of neutrino mass generation via a seesaw mechanism and being consistent with present data~\cite{Bajc,SO10-fits}. 

The breaking of SO(10) to the SM gauge group ${\cal G}_{\rm SM} \! \equiv \! {\cal G}_{321}\! =\! {\rm SU(3)_C \times SU(2)_L \times U(1)_Y}$ can be triggered in several ways, but we will be only interested in two patterns that involve one intermediate gauge group at a high scale $M_I$: the Pati-Salam (PS) group \cite{PS} ${\cal G}_{422}\!=\!{\rm SU(4)_C \times SU(2)_L \times SU(2)_R}$ and the minimal left-right (LR) symmetry group \cite{LR} ${\cal G}_{3221}\!=\!{\rm SU(3)_C \times SU(2)_L \times SU(2)_R \times U(1)_{B-L}}$. To achieve the desired symmetry-breaking in these two scenarios, one would necessarily need to introduce scalar multiplets that acquire vacuum expectation values (vevs) at the corresponding high scales. 

For the Pati-Salam scenario, the breaking chain from SO(10) to the SM gauge group is {\it e.g.} achieved by the  ${\bf (15, 1, 1)}$ component of the scalar representation $\mathbf{210_H}$ which acquires a vev at the GUT scale $M_U$, and by the $\mathbf{\overline{126}_H}$ that acquires a vev at the intermediate scale $M_I$. In turn, in the minimal left-right scenario, the symmetry should be broken first by the $\mathbf{45_H}$ which acquires a vev at the GUT scale and then by the $\mathbf{ \overline{126}_H}$ that acquires it at the intermediate scale \cite{Babu:2016bmy}. One thus has 
\begin{eqnarray}
&& \text{PS}:  \quad \text{SO(10)}|_{M_{U} } \xrightarrow{\langle \mathbf{210_H} \rangle} {\cal G}_{422}|_{M_{I} } \xrightarrow{\langle \mathbf{\overline{126}_H}\rangle
} {\cal G}_{321}|_{M_Z}\xrightarrow{\langle \mathbf{10_H} \rangle} {\cal G}_{31}  \, ; \\
&& \text{LR}: \quad \text{SO(10)}|_{M_{U}} \xrightarrow{\langle \mathbf{45_H} \rangle} {\cal G}_{3221}|_{M_{I}} \xrightarrow{\langle \mathbf{\overline{126}_H} \rangle} {\cal G}_{321}|_{M_Z} \xrightarrow{\langle \mathbf{10_H} \rangle} {\cal G}_{31} \, .
\label{breakingchain}
\end{eqnarray}

According to the extended survival hypothesis~\cite{Survival}, all the scalar fields that do not participate in the symmetry breaking patterns above by acquiring vevs will have masses of the order of the high scales $M_U$ and $M_I$. In these two breaking chains, the scalar content that acquires vevs at the intermediate scale $M_I$ or at the electroweak scale $M_Z$ consists of, respectively, the $\mathbf{\overline{126}_H}$ and $\mathbf{10_H}$ representations. More specifically, of the SO(10) scalar representations that can be decomposed under the intermediate gauge groups, only certain scalar fields from $\mathbf{10_H}$ and $\mathbf{\overline{126}_H}$ have masses below the GUT scale  and will contribute to the renormalisation group equations (RGEs) between the two scales $M_I$ and $M_U$. These are, in the PS scenario, ${\bf (1, 2, 2)}$ ($\Phi_{10}$) from $\mathbf{10_H}$ and ${\bf (15, 2, 2)} \oplus {\bf (10, 1, 3)}$ ($\Sigma_{126}\oplus \Delta_R $) from $\mathbf{ \overline{126}_H}$ and, in the minimal LR scenario,  ${\bf (1, 2, 2, 0)}$ from $\mathbf{10_H}$ and the ${\bf (1, 2, 2, 0)} \oplus {\bf (1, 1, 3, 2)}$ from $\mathbf{ \overline{126}_H}$. We will thus only consider this restricted set of scalar fields between the GUT and intermediate scales\footnote{This in contrast to most studies which are done in this context as,  generally,  a very complicated scalar sector of the SO(10) group is needed to fit the low energy spectrum, in particular the fermion (including the light and sometimes even the heavy neutrino sector) masses and mixings, by adjusting the numerous input parameters that are available.}.

At low energies, among the two Higgs bi-doublets that we had at a high energy scale, only two Higgs doublets survive and develop vevs at the electroweak scale. Thus, in our study, we will  have in fact a model with two Higgs doublet fields $H_u$ and $H_d$ that couple separately to isospin $+\frac12$ and $-\frac12$ fermions and acquire vevs $v_u$ and $v_d$ 
\begin{eqnarray}
\left \langle H_u \right \rangle = \frac{1}{\sqrt{2}} \binom{0}{v_u} \quad , \quad \left \langle H_d \right \rangle = \frac{1}{\sqrt{2}} \binom{0}{v_d} \, ,
\label{vev1}
\end{eqnarray}
to give masses to the $W,Z$ bosons implying the relation $\sqrt{v_u^2+v_d^2}\!=\!v_{\rm SM} \! \simeq \! 246 \, \text{GeV}$;  we then define the ratio of these two vevs to be $\tan \beta =v_u/v_d$. The most general renormalizable scalar potential of this two Higgs doublet model can be found in Ref.~\cite{Branco:2011iw} to which we refer for all details. The Yukawa interactions of the fermions are those of a type-II 2HDM with a Lagrangian given by
\begin{eqnarray}
-{\cal L}_{Y}^{\text{2HDM}}= Y_{u}\bar{Q}_L H_u \; u_R +Y_{d}\bar{Q}_L H_d \; d_R + Y_{e}\bar{L}_L {H_d}\; e_R + {\rm h.c.} \, ,
\label{eq:Yuk-2HDM}
\end{eqnarray}
with $Q_L/L_L$ the quark/lepton left-handed doublets and $f_R$ the right-handed singlets. In our study, only the third generation of fermions will be considered and the small Yukawa couplings of the first two generations will be neglected.

At the intermediate scale $M_I$, the minimal Yukawa interaction Lagrangian is obtained when only two Higgs bi-doublets couple to fermions. One of them should be  from the $\mathbf{ \overline{126}_H}$ which also has a triplet field that breaks the left-right symmetry. The other can be chosen to be the $\mathbf{10_H}$. Starting from eq.~(\ref{eq:LY-general}), the Yukawa Lagrangian for fermions at the intermediate scale $M_I$ can be written in the considered two schemes as
\begin{eqnarray}
\label{eq:Yuk-PS}
-{\cal L}_Y^{PS} &=& \bar{F}_L(Y^{10}_{PS} \Phi_{10}+Y^{126}_{PS} \Sigma_{126})F_R + F_R^T Y^R_{PS}  C \overline{\Delta_R} F_R  +{\rm h.c.} \, , \\
-{\cal L}_Y^{LR} &=& \bar{Q}_L(Y^{10}_{LR}\Phi_{10}+Y^{126}_{LR}\Sigma_{126})Q_R + \bar{L}_L(Y^{10}_{LR}\Phi_{10}+Y^{126}_{LR}\Sigma_{126})L_R \nonumber\\
&+&  {\small \frac12} L_R ^T Y^R_{LR} i \sigma_2 \Delta_R L_R  +{\rm h.c.} \, , 
\label{eq:Yuk-LR}
\end{eqnarray}
where $F_{L,R}$ are generic left or right-handed quark/lepton fields and $\sigma_2$ a Pauli matrix. In both cases, we have assumed that terms like $\bar{F}_L^T \Tilde{\phi} F_R$ with $\phi=\Phi$ or $\Sigma$ and $\Tilde{\phi}=  \sigma_2^T \phi^* \sigma_2$ are forbidden by suitably chosen ${\rm U(1)_Y}$ charges~\cite{Fukuyama:2002vv}. Below the intermediate scale, the PS and LR models include, besides the triplet field $\Delta_R$ that gives masses to the heavy neutrino species, four Higgs doublets: two doublets $\phi_1$ and $\phi_3$ with opposite hypercharge from the $(\mathbf{1},\mathbf{2},\mathbf{2})$ representation and the doublets $\phi_2$ and $\phi_4$ again with opposite hypercharge from $(\mathbf{15},\mathbf{2},\mathbf{2})$. The fields $\phi_1$ and $\phi_2$ couple to up-type quarks and heavy neutrinos, while $\phi_3$ and $\phi_4$ couple to down-type quarks and the light leptons. 

While the triplet fields acquire a very large vev, $ \langle \Delta_R \rangle = v_R \sim {\cal O}(M_I)$,  the bi-doublet fields acquire vevs of the order of the electroweak scale which implies that $\sum_{i=1}^4 v_i^2 = v_{\rm SM}^2$. This ensures that the right-handed gauge bosons are very heavy,  $M_{W_R}, M_{Z_R} \approx g v_{R}$, while the ${\rm SU(2)_L}$ $W$ and $Z$ bosons have weak scale  masses, $M_{W}, M_{Z} \approx g v_{\rm SM}$. In fact, only two linear combinations of the four scalar doublet fields $\phi_1 \cdots \phi_4$ will have weak scale masses, while the two other field combinations will have masses close to the very high scale. One has thus to tune the scalar potentials of the two scenarios to achieve this situation and discussions about the constraints to which it leads can be found in Refs.~\cite{Deshpande:1990ip,Dev:2018foq} for instance. The two fields with weak scale masses will be ultimately identified with the doublets $H_u$ and $H_d$ of our low energy 2HDM.  At the intermediate scale $M_I$, these fields should match the $\Phi_{10}$ and $\Sigma_{126}$ fields, the interactions of which have been  given in eqs.~(\ref{eq:Yuk-PS},\ref{eq:Yuk-LR}) as will be discussed shortly. 

\subsection*{3 Gauge coupling unification}

Assuming the 2HDM structure at low energies and the two breaking patterns of SO(10) down to the SM group with the intermediate scale $M_I$ discussed previously, namely PS and LR , we study the renormalisation group running of the three SM gauge couplings  $\alpha_i = g_i^2/(4\pi)$. We closely follow Ref.~\cite{Meloni:2019jcf} in which the standard case with only one electroweak Higgs doublet was studied. The analytical expressions for the gauge coupling RGEs at the two loop level, including the relevant $\beta$ functions can be found, {\it e.g.,} in Ref.~\cite{two-loop-RGEs} where the dependence of the number of Higgs doublets is explicitly given. Naively, the more intermediate scale scalar particles are included in the running of the couplings, the lower would be the resulting unification scale.

At the intermediate scale, threshold effects \cite{Threshold} due to all the particles that have masses in the vicinity of $M_I$, and in particular all the scalar fields that develop vevs at this scale, will be active. These higher order corrections will modify the matching conditions of the gauge couplings at the scale of symmetry breaking, depending on the particle content. For a symmetry breaking from a group ${\cal G}$ to a subgroup ${\cal H}$ at the scale $\mu$, the matching conditions with the threshold corrections take the form
\begin{eqnarray}
\alpha_{i,{\cal H}} ^{-1} (\mu )= \alpha_{i,{\cal G}} ^{-1} (\mu ) - {\lambda_{i, {\cal H}} ^{\cal G}}/{(12 \pi)} \, ,
\label{eq:threshold}
\end{eqnarray}
where $i=1,2,3,...$ refers to the particular gauge coupling $\alpha_i$ and $\lambda_{i, {\cal H}} ^{\cal G}$ are usually weighted by the parameters $\eta_i=\ln ({M_{i}}/{\mu})$ with $M_{i}$ being the masses of the heavy particles integrated out at the low energy. The complete expressions for the one-loop threshold corrections $\lambda_{i, {\cal H}} ^{\cal G}$ are given in Ref.~\cite{Meloni:2019jcf} (see tables IV and VI of their Appendices B and C, respectively). As a result, the intermediate and the unification scales $M_I$ and $M_U$ could be shifted by an order of magnitude or more even when only small threshold corrections are included. In the following, we show an explicit example of gauge coupling unification in the PS and LR breaking chains when these thresholds are included.

We start with the following initial conditions for the SM gauge couplings calculated in the $\overline{\text{MS}}$ renormalization scheme with two-loop accuracy and first evaluated at the electroweak scale that we take to be the $Z$ boson, mass $M_Z=91.2$ GeV~\cite{PDG},
\begin{eqnarray}
[g_Y(M_Z), g_{2}(M_Z), g_3(M_Z)] = [0.3574, 0.6517, 1.2182] \, ,
\label{eq:gis}
\end{eqnarray}
where $g_Y$ should be normalized with the usual GUT condition leading to $\alpha_1 / \alpha_Y=5/3 $. Using the two-loop RGEs in the case in which two Higgs doublets are present at low energies and including the relevant threshold corrections following Ref.~\cite{Meloni:2019jcf}, we determine the point at which the couplings intersect when appropriately adjusting the intermediate scale $M_I$. In the two symmetry breaking chains that we consider, the tree-level matching conditions that determine the gauge couplings of the intermediate scale models from the low energy ones read
\begin{eqnarray}
{\rm PS}: & 
\alpha_4^{-1} (M_I) = \alpha_3^{-1} (M_I) \, , \ 
\alpha_{2L}^{-1} (M_I) = \alpha_2^{-1} (M_I)  \, , \ 
\alpha_{2R}^{-1} (M_I) =\frac53 \alpha_Y^{-1} (M_I)-\frac23 \alpha_3^{-1} (M_I)  \, ,   ~~\nonumber \\ 
{\rm LR}: &\!\!  \hspace*{-5mm}
\alpha_3^{-1} (M_I) = 
\alpha_{2L}^{-1} (M_I) = \alpha_3^{-1} (M_I) \, , \ 
 \alpha_{B\!-\!L}^{-1} (M_I) = \kappa \alpha_{2R}^{-1} (M_I) =
 \big( \frac{2\kappa+ 3}{5} \big)^{-1} \alpha_Y^{-1} (M_I)  \, , 
\label{eq:match-MI}
\end{eqnarray}
where in LR we assume $\alpha_{B\!-\!L}^{-1} (M_I) = \kappa \alpha_{2R}^{-1} (M_I)$ as we are matching three couplings to four; this normalization factor $\kappa$ of ${\cal O}(1)$ is to be solved with the scales $M_I$ and $M_U$.

Note that in eq.~(\ref{eq:gis}), we have ignored, for simplicity, the experimental errors on the couplings constants (as well as the theoretical uncertainties) and kept only the central values. These errors,  the largest of which being the one that affects the strong coupling constant $\alpha_3$ which is at the percent level,  will generate an uncertainty on the derived GUT an intermediate scales of the order of a few percent only and, hence, do not affect our discussion in a significant way as will be shown shortly. 

In our analysis, the RGEs of the gauge couplings are solved up to two loop order with the help of the program SARAH~\cite{SARAH}, and the inclusion of the thresholds corrections is performed by randomly sampling parameters $\eta_i = \ln(M_i/\mu)$ within the range of $\eta_i \in [-1,1]$. We then impose the tree-level matching conditions of the gauge couplings at the intermediate  scale as in eq.~(\ref{eq:match-MI}) when the one-loop threshold corrections are included as in eq.~(\ref{eq:threshold}), and determine the values of the two scales $M_I$ and $M_U$ for each sampling parameter set. More precisely, we take at least 10,000 points for the parameters $\eta_i$ within the range of $\eta_i \in [-1,1]$ and determine the sets of all scales that allow for gauge coupling unification. 

In the two intermediate SO(10) scenarios that we consider, gauge coupling unification with the inclusion of threshold corrections can, for instance, be achieved for the following values of the unification and intermediate scales 
\begin{eqnarray}
\label{eq:scales}
{\rm PS}: && M_U=7.5\times 10^{15}~{\rm GeV~~and~~}M_I=6.6 \times 10^{10}~\text{GeV}\, , 
\nonumber \\
{\rm LR}: && M_U=3.9\times 10^{15}~{\rm GeV~~and~~}M_I=6.0 \times 10^{9}~\text{GeV}\, .
\end{eqnarray}

\begin{figure}[!t]
\begin{center}
\hspace{-1cm}
\mbox{
\includegraphics[width=7.5cm,clip]{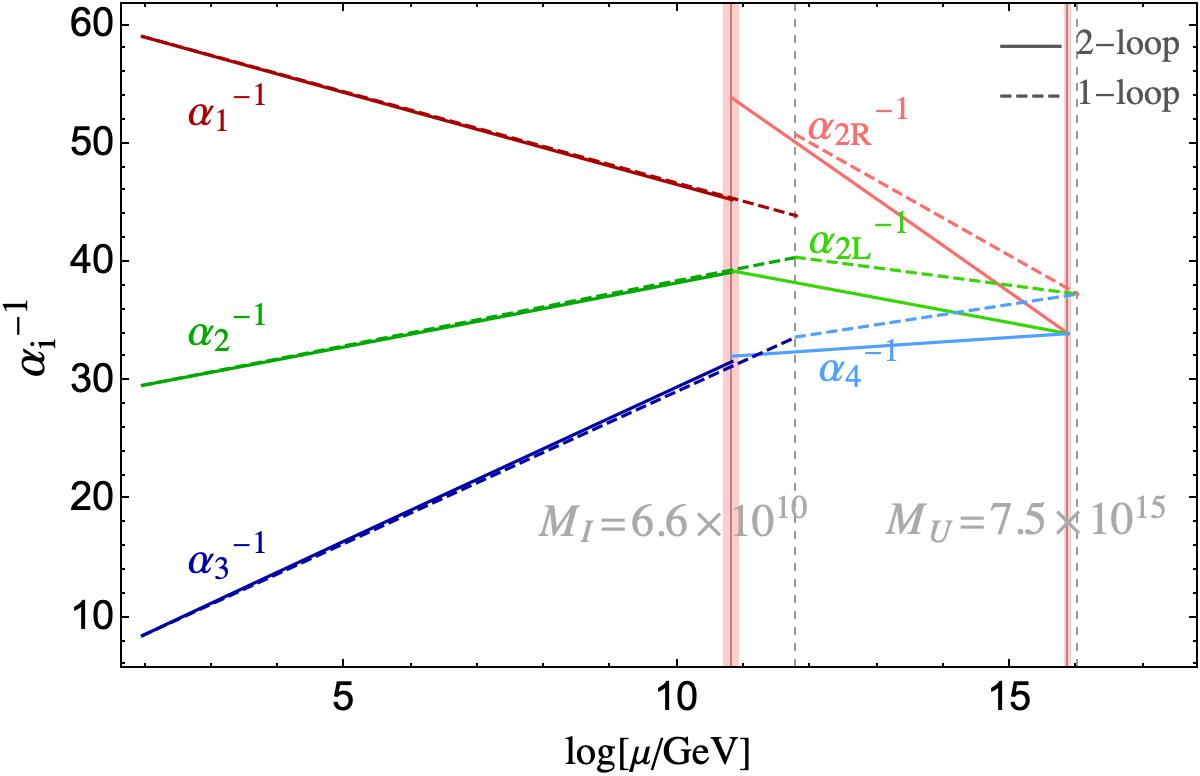} \hspace*{0.cm} 
\includegraphics[width=7.5cm,clip]{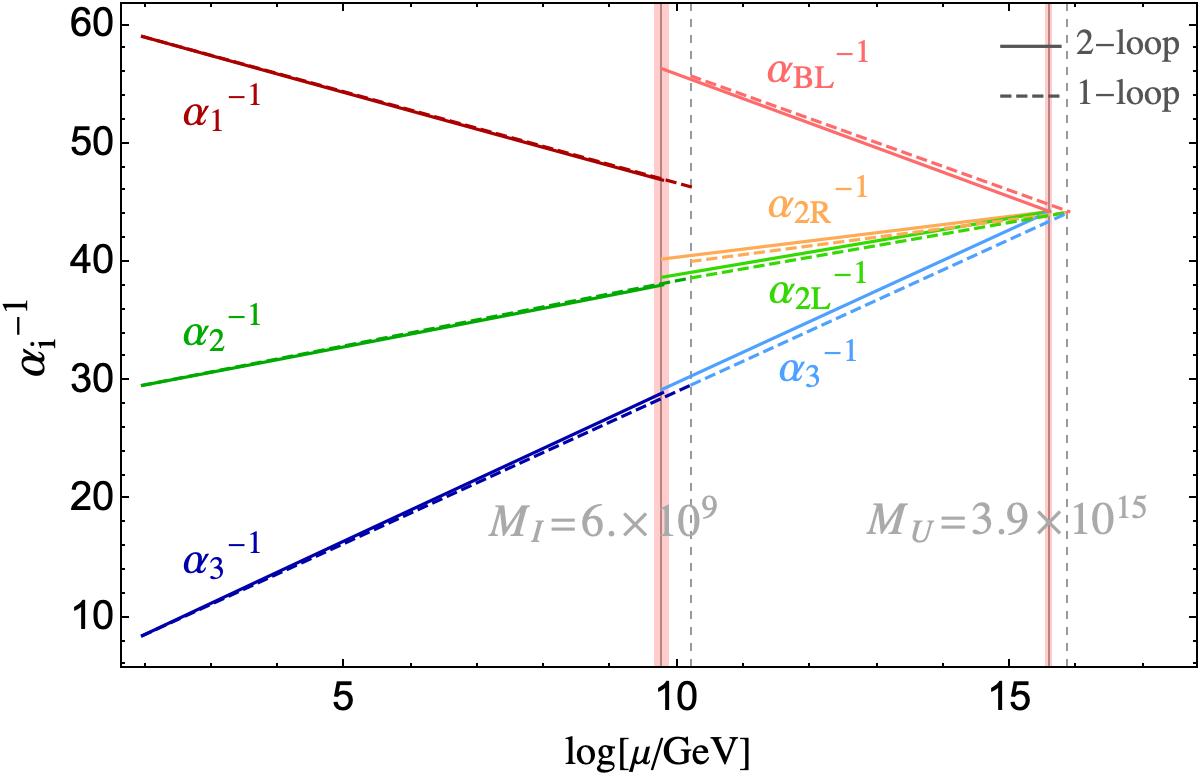}
\hspace*{-0.5cm} 
}
\end{center}
\vspace*{-5mm}
\caption{The evolution of the inverse of the gauge coupling constants $\alpha_i = g_i^2/(4\pi)$ as a function of the energy scale $\mu$ in the 2HDM+${\cal G}_{422}$ Pati-Salam model (left) and 2HDM+${\cal G}_{3221}$ minimal LR model (right) at the one-loop (dashed lines) and two-loop (solid lines) orders. The GUT and intermediate scales $M_U$ and $M_I$ are indicated and the threshold effects are included. The red bands reveal the effects of the experimental uncertainties in the measurement of the couplings
}
\label{fig:gauge-unif}
\vspace*{-5mm}
\end{figure}

The evolution of the inverse of the coupling constants $\alpha_i^{-1}$ from the scale $M_U$ down to $M_I$ and then down to $M_Z$ is shown in Fig.~\ref{fig:gauge-unif} as a function of the energy scale in the two breaking patterns PS (left panel) and LR (right panel) when the two-loop (solid lines) and one-loop (dashed lines) RGEs are used and the threshold effects are included at the intermediate scale.  While the three couplings are clearly different at the scale $M_I$ of the order of a few times $10^{10}$ GeV (as required to reproduce neutrino phenomenology), the slope are significantly modified at this energy by the additional contributions so that the couplings meet at a scale $M_U$ of the order of a few times $10^{15}$ GeV (which is high enough to prevent fast proton decay). Both the two-loop corrections and the threshold corrections have a noticeable impact and make the intermediate scale lower. The small impact of the experimental errors on the couplings is illustrated by the narrow red bands at the scales $M_I$ and $M_U$.

\subsection*{4 Yukawa coupling unification}
 
We now turn to the Yukawa sector of the theory. As already mentioned, we will ignore the very small Yukawa couplings of the first and second generation fermions\footnote{As in the supersymmetric case, fermions with masses below a few GeV cannot be realistically described in our approach as it will be plagued by strong interaction uncertainties when running the RGEs down to the fermion mass scale.}  and consider only those  of the top quark, the bottom quark and the tau lepton, neglecting all possible mixings. Below the intermediate scale $M_I$, the Yukawa interactions of these fermions are those of a type-II 2HDM with a Lagrangian  given by eq.~(\ref{eq:Yuk-2HDM}). It leads to the following relations between the fermion masses and the Yukawa couplings
\begin{equation}
m_t =\frac{1}{\sqrt{2}}Y_{t} v_{u} \, , \  \ \
m_b =\frac{1}{\sqrt{2}}Y_{b} v_{d} \, , \ \  \
m_\tau =\frac{1}{\sqrt{2}}Y_{\tau} v_{d} \, .
\label{eq:fmasses}
\end{equation}

In the region between the intermediate scale and the GUT scale, we assume the Yukawa structure of eqs.~(\ref{eq:Yuk-PS}) and (\ref{eq:Yuk-LR}) for the PS and minimal LR breaking patterns, respectively. With a real $\mathbf{10_H}$ representation with its vevs denoted by $v_{10}^u=v^{d \, *}_{10}=v_{10}$  and by adopting a phase convention in which $v_{10}$ is real (this can be done via, {\it e.g.,} an SU(2) rotation)~\cite{Babu:2016bmy}, and denoting by $v_{126}^{u,d}$  the vevs of the $\Sigma_{126}$ field, the fermion masses for the two considered breaking chains will be given by
\begin{eqnarray}
m_t \!=\! \frac{v_{10} Y_{10}^{PS} \!+ \! v_{126}^u Y_{126}^{PS} } {\sqrt 2 }  , \,
m_b \!=\! \frac{v_{10} Y_{10}^{PS}\!+\! v_{126}^d Y_{126}^{PS}    }{\sqrt 2 }   , \,
m_\tau\! = \! \frac{ v_{10} Y_{10}^{PS} \!-\!3 v_{126}^d Y_{126}^{PS} }{\sqrt 2 } ,~~ 
 \nonumber \\
m_t \!= \! \frac{v_{10} Y_{10,q}^{LR} \!+\! v_{126}^u Y_{126,q}^{LR} }{\sqrt 2 }, \,
m_b \!=\! \frac{v_{10} Y_{10,q}^{LR}\!+\! v_{126}^d Y_{126,q}^{LR} }{\sqrt 2 }, \, 
m_\tau \!=\! \frac{v_{10} Y_{10,l}^{LR} \!+\! v_{126}^d Y_{126,l}^{LR} }{\sqrt 2 } \, .
\label{eq:fmasses2}
\end{eqnarray}

Finally, above the GUT scale $M_U$, the third generation Yukawa couplings are unified as in eq.~(\ref{eq:LY-general}) and are given by  
\begin{eqnarray}
m_t =v_{10} Y_{10} + v_{126}^u Y_{126}\, , \ 
m_b = v_{10} Y_{10}+v_{126}^d Y_{126} \, ,  \
m_\tau = v_{10} Y_{10} -3 v_{126}^d Y_{126} , 
\label{eq:fmasses3}
\end{eqnarray}
(with the additional masses for neutrinos $M_{\nu_D} \!= \! v_{10} Y_{10} \! - \! 3 v_{126}^u Y_{126}$ and $M_{\nu_{R}} \!= \! v_{R}Y_{126}$). The normalization factors can be absorbed into the redefinition of the Yukawa couplings at the GUT scale. The factors of 3 and the relative signs between the various terms are due to the Clebsh-Gordan coefficients coming from the vev of the traceless adjoint 15 of $\rm SU(4)$ in ${\bf (2,2,15)}$.

As the evolution of the couplings near the scale $M_I$ should be affected by threshold corrections, one should expect a significant discontinuity of the Yukawa couplings when the contributions of the numerous scalar and vector fields are included in the RGEs. Nevertheless, as these Yukawa couplings are directly related to the masses of the fermions, one can simply assume that the physical fermion masses are continuous at the scale $M_I$~\cite{Meloni:2016rnt} when these threshold corrections are included. This means that the masses calculated in the low-energy 2HDM should coincide with those obtained from the intermediate left-right or Pati--Salam models or the unified SO(10) model, up to their running. One can then consider this relation as the matching conditions for the Yukawa couplings at the intermediate and the GUT scales. For example, at the scale $M_I$, equating eqs.~(\ref{eq:fmasses}-\ref{eq:fmasses2}) for the PS and LR breaking chains leads to\footnote{For this exploratory work, we simply follow Ref.~\cite{Meloni:2016rnt} and ignore the small running of the vevs. This issue, together with other refinements,  will be postponed to a forthcoming publication.}
\begin{eqnarray}
Y_t (M_I) &\!= Y_{10}^{PS}(M_I) \frac{v_{10}}{v_u}  \!+ \! Y_{126}^{PS} (M_I) \frac{v_{126}^u}{v_u} & {\rm or}~~ Y_{10,q}^{LR}(M_I) \frac{v_{10}}{v_u}  \!+ \! Y_{126,q}^{LR} (M_I) \frac{v_{126}^u}{v_u}   , \, \nonumber \\
Y_b (M_I) &\!=  Y_{10}^{PS} (M_I) \frac{v_{10}}{v_d} \!+\! Y_{126}^{PS} (M_I) \frac{v_{126}^d}{v_d} & {\rm or}~~~ Y_{10,q}^{LR} (M_I) \frac{v_{10}}{v_d} \!+\! Y_{126,q}^{PS} (M_I) \frac{v_{126}^d}{v_d}      , \, \nonumber \\
Y_\tau (M_I) &\!= Y_{10}^{PS} (M_I) \frac{v_{10}}{v_d} \!-3  Y_{126}^{PS} (M_I) \frac{v_{126}^d}{v_d} & {\rm or}~~~ Y_{10,l}^{LR} (M_I) \frac{v_{10}}{v_d} \!+  Y_{126,l}^{LR} (M_I) \frac{v_{126}^d}{v_d} . \, 
\label{eq:match1}
\end{eqnarray}

As for the matching conditions at the GUT scale, one has to carefully take the Clebsh-Gordan factors into account for the Yukawa interactions when the field representations are embedded into the SO(10) group~\cite{Aulakh:2002zr,Fukuyama:2002vv}. One can then enforce Yukawa coupling unification by requiring the matching conditions at the GUT scale to be
\begin{eqnarray}
Y_f(M_U) & \equiv & Y_{10}^{PS}(M_U)=\frac{1}{4} Y_{126}^{PS}(M_U)    \, ,  \\
Y_f(M_U) & \equiv & Y_{10,q}^{LR} (M_U)= \frac{1}{4}Y_{126,q}^{LR}(M_U)=Y_{10,l}^{LR} (M_U)=- \frac{1}{12} Y_{126,l}^{LR} (M_U) \, , 
\label{eq:match2}
\end{eqnarray}
where the unified Yukawa coupling $Y_f(M_U)$ is taken to be a free parameter of SO(10).

One has then to fit these parameters with the actual observables, namely the top, bottom and tau masses using the relations in eq.~(\ref{eq:fmasses}) at the low energy scale, chosen again to be $M_Z=91.2$ GeV. We use the following input $\overline{\text{MS}}$ running fermion masses in the SM \cite{PDG,Bednyakov:2016onn} (we again ignore the related experimental uncertainties for now),
\begin{eqnarray}
\label{eq:exp-masses}
[m_t(M_Z), m_b(M_Z), m_\tau (M_Z)] = [168.3, 2.87,  1.73]~{\rm GeV}\, ,
\end{eqnarray}
and we then turn them into the corresponding input masses in the 2HDM by using the appropriate RGEs in the evolution from the scale of the fermion masses to $M_Z$.

In the PS model, the colored-quarks and leptons are charged under the same local SU(4) symmetry so that all fermions can be unified into the same representation $F_{L,R}$. When these fermions couple to the Higgs fields ${\bf 10_H}$, one cannot distinguish the bottom quark from the tau lepton and one should have $m_b=m_{\tau}$ if the vev $v_{126}^d$ is small, $v_{126}^d/v_d \ll 1$, as can be seen from eq.~(\ref{eq:fmasses2}). If this mass equality is still valid slightly below the intermediate scale, we should then have $Y_b(M_I)=Y_{\tau}(M_I)$ in our low energy 2HDM by virtue of eqs.~(\ref{eq:match1}) and (\ref{eq:fmasses}).  The scale at which the bottom and tau Yukawa couplings are equal, that we denote by $M_{b\tau}$, is simply determined (within some accuracy) by the point at which the curves for their RG running from the weak scale $M_Z$ upwards intersect, which critically depends on the value of the parameter $\tan\beta$.  In order to use the matching conditions given by the equations above, the scale for $b$-$\tau$ unification should be identical to the intermediate PS breaking scale,  $M_{b\tau}=M_I$, and this can be achieved by selecting the appropriate value of $\tan\beta$. 

The RGEs for the Yukawa couplings from $M_U$ to the intermediate scale $M_I$ and from $M_I$ to the electroweak scale $M_Z$ up to the two-loop level have been given in Ref.~\cite{Machacek:1983fi} in the standard case with one electroweak Higgs doublet only. In our case, we also include the additional contributions of the extra Higgs doublet at the low scale. We solve the system using again the program SARAH \cite{SARAH}.

\begin{figure}[t]
\begin{center}
\mbox{
\includegraphics[width=7.5cm,clip]{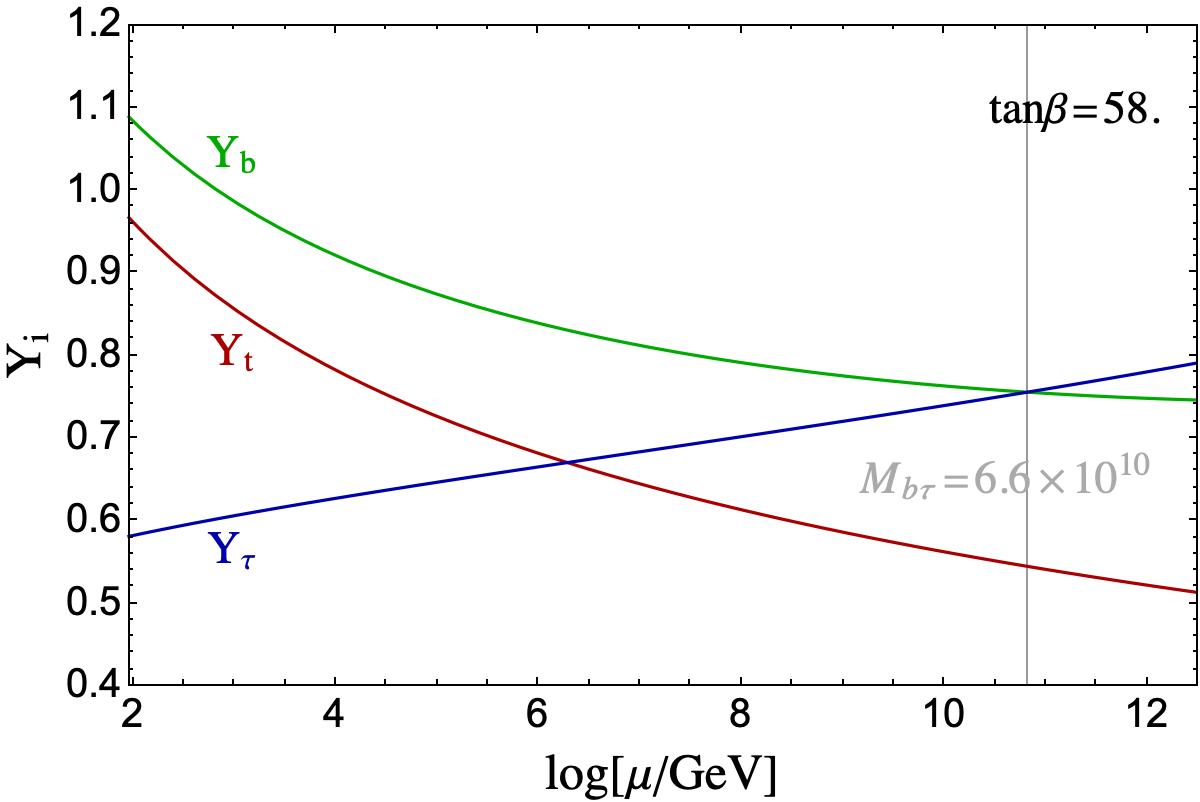}~~
\includegraphics[width=7.5cm,clip]{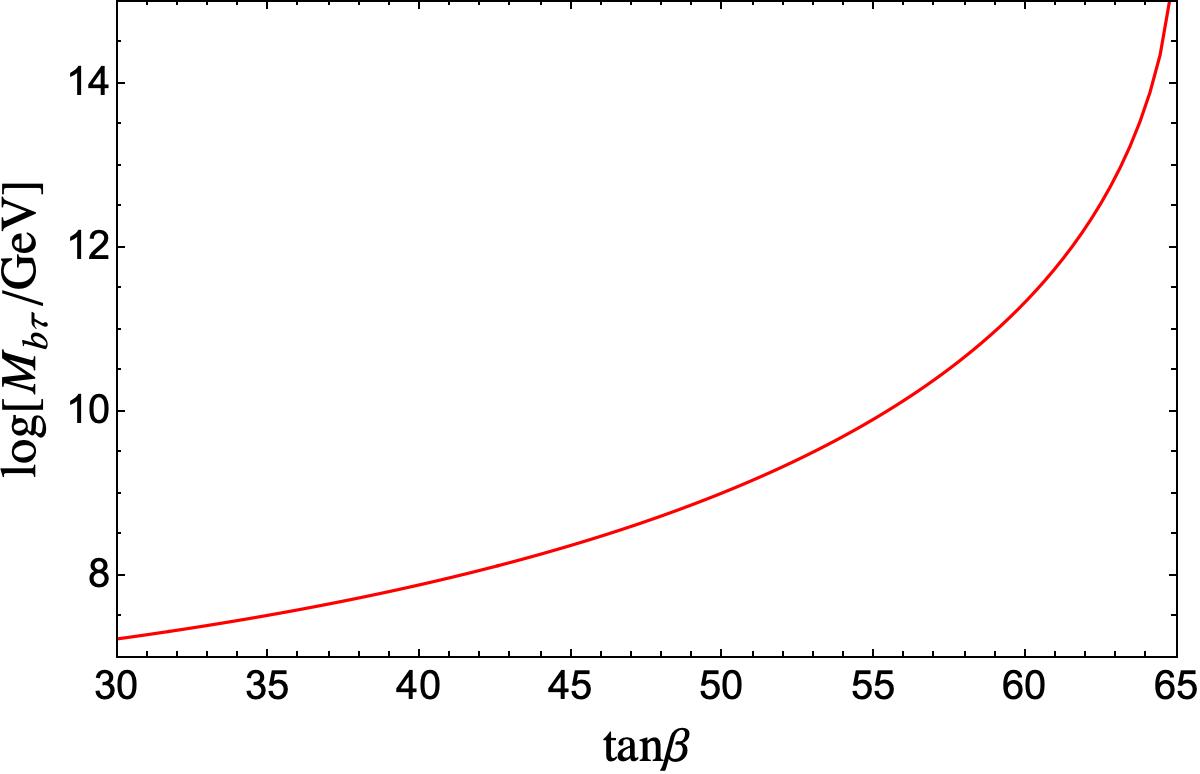} 
}
\end{center}
\vspace*{-5mm}
\caption{An example of the running of third generation fermion Yukawa couplings from the weak to the high scales for the value $\tan\beta=58$ for which the bottom and tau couplings unify at a scale $M_{b\tau}=6.6\times 10^{10}$ (left) and the dependence of this unification scale $M_{b\tau}$ on the value of  $\tan \beta$ (right).} 
\label{fig:Mbtau}
\end{figure}

In the PS model, the running of the third generation Yukawa couplings from the low to the high energy scales are shown in the left panel of Fig.~\ref{fig:Mbtau}, for the specific case where the input value $\tan \beta=58$ is chosen. One can see that, indeed, the curves for $Y_b$ and $Y_\tau$ intersect at an energy scale $M_{b\tau}\! \simeq \!7  \!\times \! 10^{10}$ GeV, which is very close to the intermediate scale for which the gauge couplings unify in the PS scheme. 

The right panel of the figure shows the dependence of the bottom-tau unification scale $M_{b\tau}$ on the ratio of vevs $\tan \beta$ and, as can be seen, intermediate scale values between $M_I=10^9$ GeV and $M_I=10^{11}$ GeV would imply high values of $\tan\beta$, in the range $\tan\beta \approx 50\!-\!60$. Note that the value of $\tan \beta$ cannot be arbitrarily high, $\tan \beta$ \raisebox{-0.13cm}{~\shortstack{$<$ \\[-0.07cm] $\sim$}}~70 in the specific cases we are discussing here, in order to avoid that the Yukawa couplings run to non-perturbative values at these scales.  

In the minimal LR model, the discussion above does not hold and the bottom and tau Yukawa couplings do not unify at the intermediate scale. Nevertheless, one should have close if not equal values for the Yukawa terms $Y_{10,q}^{LR}$ and $Y_{10,l}^{LR}$ such that they can run to a common value at the scale $M_U$ where, according to eq.~(18), one has $Y_{10,q}^{LR} = Y_{10,l}^{LR}$. 

We come now to the unification of all Yukawa couplings. With the four Yukawa couplings, the randomly chosen one $Y_f(M_U)$ and the three weak scale ones $Y_t,Y_b,Y_\tau$, the scales $M_U$ and $M_I$ to be determined from gauge coupling unification, five parameters are needed to entirely describe our Yukawa sector: the 2HDM vevs $v_u, v_d$ at $M_Z$ and the vevs $v_{126}^u, v_{126}^d$ and $v_{10}$ of the bi-doublets at the scale $M_I$. Nevertheless, we have many constraints at hand as, besides the matching relations given in eqs.~(\ref{eq:match1}-\ref{eq:match2}), one needs to reproduce the experimental values of the standard particle masses.

Indeed, at both the scales $M_Z$ and $M_I$, the correct $W,Z$ masses should be reproduced, giving  $\sqrt{ v_u^2\!+\!v_d^2}\!=\!v_{\rm SM}\!=\!246 \, {\rm GeV} \!=\! v_{10}^2\!+\! v_{126}^2 \!+ \! v_{126d}^2$\cite{Deshpande:1990ip}. One needs also to reproduce the heavy fermion masses at the weak scale $M_Z$, eq.~(\ref{eq:exp-masses}), using the relations of eq.~(\ref{eq:fmasses}). We will assume that there is an uncertainty of the order of $2\%$ in reproducing all these particle masses.  This uncertainty, which is sufficiently small for our purpose (and allows us to have some solutions for the coupled RGE's), is introduced not only because of the experimental errors (e.g. on $\alpha_3$ and the top and bottom masses) but also the theoretical ones from various sources such as the higher order effects in the RGEs, the higher order threshold corrections, the possible running of the vev's,  etc.

\begin{figure}[!h]
\begin{center}
\includegraphics[width=11.cm,clip]{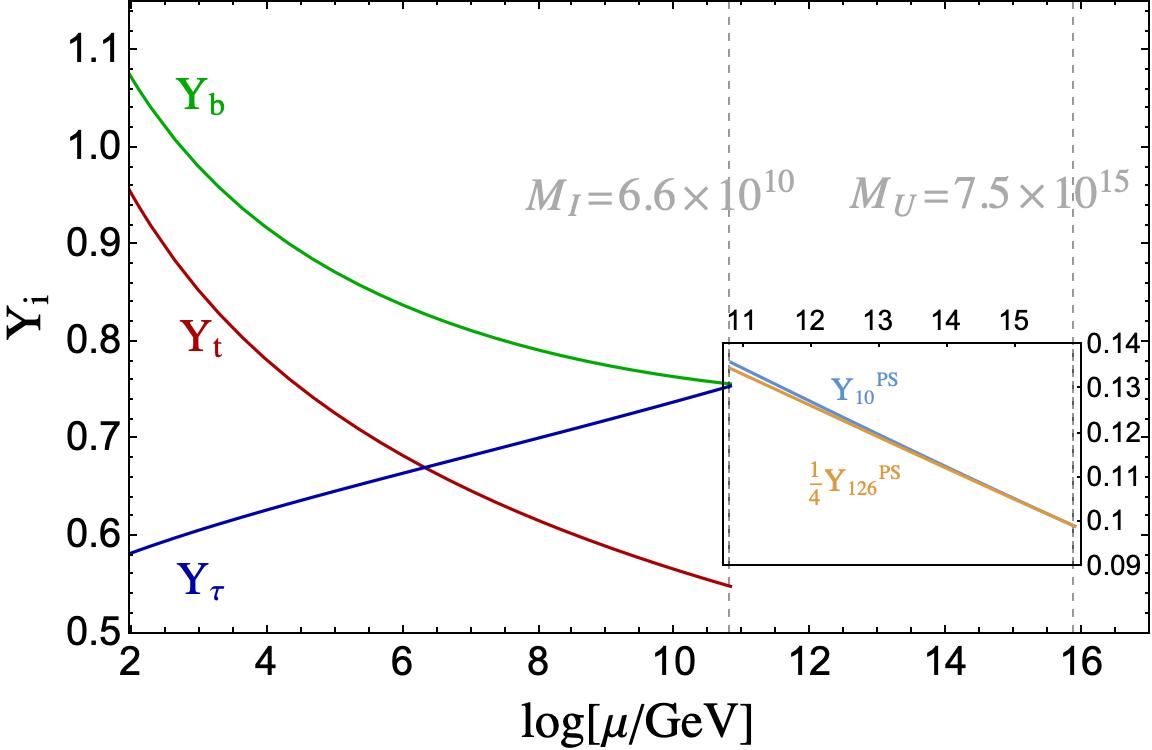}\\[2mm]
\includegraphics[width=11.cm,clip]{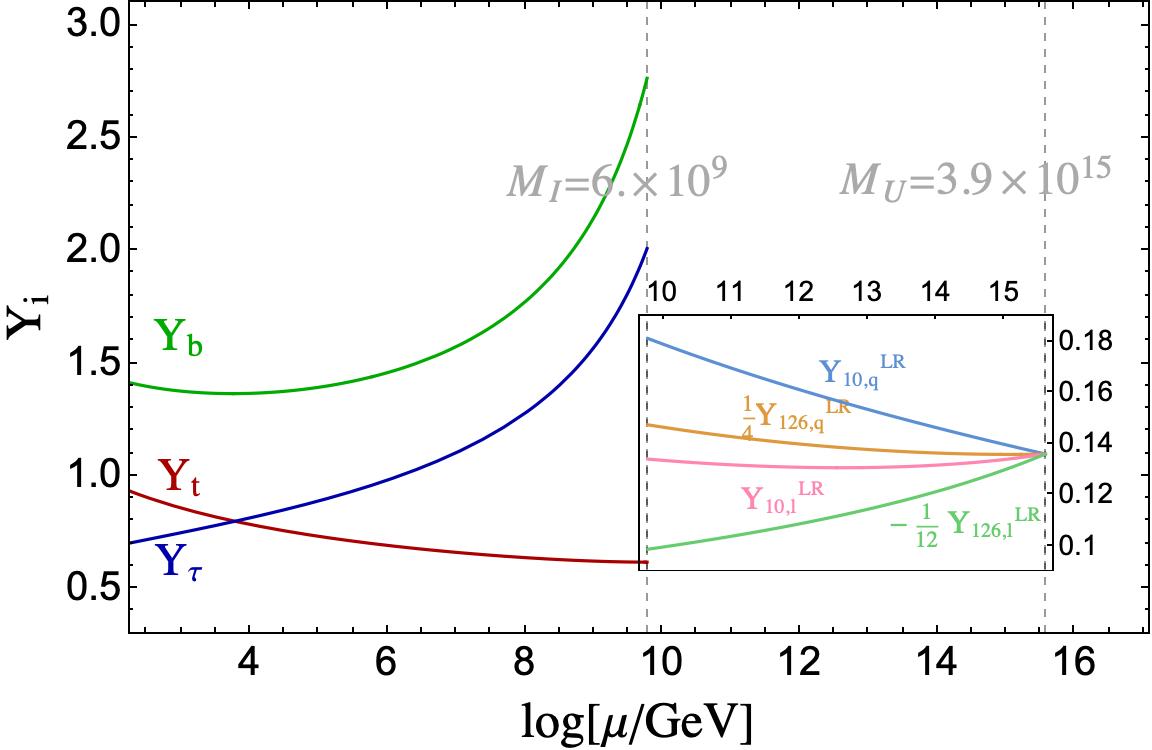}
\end{center}
\vspace*{-5mm}
\caption{The renormalisation group running of the Yukawa couplings at two-loop  in the 2HDM+${\cal G}_{422}$ model (top) and 2HDM+${\cal G}_{3221}$ model (bottom) including matching conditions and  threshold effects at an intermediate scale $M_I$ for a parameter set that is compatible with the observed top, bottom, tau as well as gauge boson masses.}
\label{fig:Yukawa-run}
\vspace*{-2mm}
\end{figure}

For completeness, we make sure in addition that the electroweak vacuum remains stable up to the intermediate scale $M_I \approx 10^{10}$ GeV for the chosen top quark and SM-like Higgs boson masses. To do so, we use the necessary and sufficient conditions of Refs.~\cite{Branco:2011iw,Deshpande:1977rw} on the 2HDM quartic scalar couplings to ensure the scalar potential to be bounded from below, with the input values for the relevant weak scale parameters given in Ref.~\cite{Buttazzo:2013uya}. As an additional and final constraint, we force the three Yukawa couplings to remain perturbative at all scales by imposing the conditions $Y_i^2  (\mu)/(4\pi) \leq 1$.   

We then scan this constrained parameter space in order to find a viable solution to the system of equations. Our main results are displayed in Fig.~\ref{fig:Yukawa-run} and in Table \ref{tab:Yukawas}. 

Fig.~\ref{fig:Yukawa-run} shows the evolution of the three Yukawa couplings as a function of the energy scale in the PS (upper panel) and LR (lower panel) scenarios and it can be seen that all of them reach a common value, $Y_f (M_U) \approx 0.1$, at the same GUT scale $M_U$ that leads to gauge coupling unification eq.~(\ref{eq:scales}).  At the intermediate scale $M_I$ that is also required by gauge coupling unification, eq.~(\ref{eq:scales}), one notices the discontinuity for the Yukawa couplings which is due to the matching conditions. 

Finally, in Table~\ref{tab:Yukawas}, we show examples of points in the 2HDM+PS and 2HDM+LR model parameter spaces that satisfy all the criteria discussed above and list the sets of values for the three fermion couplings and all the relevant vevs which lead to Yukawa coupling unification, with the GUT and intermediate scales that allow for gauge coupling unification, eqs.~(\ref{eq:scales}), and with all constraints implemented.

One can see that in both the PS and LR breaking schemes, one obtains approximately the same unified Yukawa coupling $Y_f (M_U) = {\cal O}(0.1) $.  One can also see that the relations $\sum_i v_i^2\!=\!v_{\rm SM}^2$ are fulfilled at the relevant scales and that all Yukawa couplings are such that their squares are smaller than $4\pi$ even at $M_I$. 

In both cases, the obtained values of the input 2HDM parameter $\tan \beta\!=\! v_u/v_d$ at the electroweak scale are large but still reasonable, approximately $\tan \beta\!=\!58$ and $\tan\beta\!=\!70$ for the PS and LR models respectively, as they ensure that the bottom quark Yukawa coupling remains perturbative at all energy scales before $M_U$, and give the correct hierarchy of quark masses at the electroweak  scale, $\tan \beta  \! \approx \! m_t/m_b$.

\begin{table}[t]
    \centering
    \begin{tabular}{|c|ccc|ccc|c||cc|ccc|}
    \hline
       scale  & & $M_Z$ & & & $M_I$ & &    $M_U$ & & $\hspace*{-.5cm}M_Z\hspace*{.5cm}$ & &   $M_I$  & \\ \hline
        & $Y_{t}$ &  $Y_{b}$ & $Y_{\tau}$ & $Y_{t}$ &  $Y_{b}$ & $Y_{\tau}$ &    $Y_f$ & $v_{d}$ & $v_{u}$ & $v_{10}$ & $v_{126}^u$ & $v_{126}^d$ \\
       \hline
       PS & 0.97 & 1.09 & 0.58 & 0.55 & 0.76 & 0.75 & 0.10 &    4.21 & 246.2 & 23.4 & 244.7 & 0.004  \\ \hline
       LR & 0.97 & 1.44 & 0.68 & 0.62 & 2.76 & 2.01 & 0.14 & 3.50 & 246.2 & 53.2 & 241.0 & 0.079  \\ \hline
    \end{tabular}
    \caption{A set of the third generation fermion Yukawa couplings at the scales $M_Z,M_I$ and $M_U$, and the relevant vevs at the weak and intermediate scales, which fit all observables within 2\% accuracy at the two-loop level and lead to both gauge coupling and Yukawa coupling unification in SO(10) with PS and LR intermediate breaking.}
    \label{tab:Yukawas}
\vspace*{-2mm}
\end{table}

Hence, Yukawa coupling unification can also be achieved in a simple manner in a non-\-supersymmetric SO(10) scenario. One can arrange to achieve it for lower values of $\tan\beta$ than above, at a minimal cost and without affecting the simplicity of the approach,  by complexifying the $\mathbf{10_H}$ representation. One still makes use of two Higgs bi-doublets above the scale $M_I$ but there are four non-zero vevs instead of three as $v_{10}^u \neq v_{10}^d$. This additional input can be adjusted to have more adequate solutions to the system of Yukawa coupling RGEs. This possibility, as well as other interesting extensions of the simple scheme proposed here, will be addressed in a forthcoming publication.

 
\subsection*{5 Conclusions}

We have analyzed the possibility of unifying the Yukawa couplings of third generation fermions in the context of a non-supersymmetric SO(10) scenario with intermediate breaking, focusing on the Pati-Salam and minimal left-right breaking chains. The framework that we adopt is rather simple as the relevant scalar sector of the theory consists of only two Higgs bi-doublets at the intermediate breaking scale, $M_I ={\cal O} (10^{10})$~GeV, reducing to a two Higgs doublet model of type II at the electroweak scale. 

We first discussed gauge coupling unification which can indeed be achieved at a GUT scale close to $M_U \approx 10^{16}$ GeV, by including the threshold effects of the scalar multiplets that appear at the intermediate scale. This is somehow expected as the contribution of the additional electroweak Higgs doublet (and all scalar fields in general) does not significantly modify the running of the gauge coupling constants. 

We have then studied the renormalisation group running of the Yukawa couplings of the top and bottom quarks and the tau lepton in the Pati-Salam and minimal left-right SO(10) scenarios, with the proper matching conditions at the unification, intermediate and electroweak scales. We have performed a scan of the parameter space of the two models, imposing that the phenomenology at low energy and, in particular the third generation fermion and the electroweak gauge boson masses, is correctly reproduced within 2\% accuracy. We find that the unification of the  Yukawa couplings of third generation can be indeed realized in  regions of the parameter space in which the ratio of the two electroweak Higgs doublet vevs is large, $\tan\beta \approx 60$.   
  
Hence, similarly to the well known and widely studied supersymmetric case, not only gauge but also Yukawa coupling unification can be achieved in SO(10) while using a rather simple Higgs sector and retaining a viable particle spectrum at the weak scale. 

An interesting feature of this possibility is that while most of the ingredients of the conventional SO(10) model are expected to be at a too high scale, ${\cal O} (10^{10})$~GeV, to be probed effectively in collider experiments, our scenario requires a second Higgs doublet at low energies. The model thus predicts additional Higgs particles  with weak scale masses which could be searched for and eventually be observed at the Large Hadron Collider or at the next generation of high-energy colliders.\bigskip  

\noindent \textbf{Acknowledgement:} We thank Paco del Aguila, Renato Fonseca, Luca Marzola and Davide Meloni for discussions and comments on the manuscript, and J\'er\'emie Quevillon for collaboration at an early stage of the study. This work was supported by the Estonian Research Council grants MOBTT86, PRG803 and by the EU through the European Regional Development Fund CoE program TK133 ``The Dark Side of the Universe." A.D. is also supported by the Junta de Andalucia through the Talentia Senior program as well as by A-FQM-211-UGR18, P18-FR-4314 with ERDF.



\end{document}